# Single Carrier Architecture for High Data Rate Wireless PAN Communications System

Lahatra Rakotondrainibe, Yvan Kokar, Gheorghe Zaharia, Guy Grunfelder, Ghaïs El Zein

Université Européenne de Bretagne, France
INSA, IETR, UMR 6164, F-35708 Rennes
Tel: (33) 2 23 23 87 13 Fax: (33) 2 23 23 84 39
lrakoton@insa-rennes.fr, yvan.kokar@insa-rennes.fr, gheorghe.zaharia@insa-rennes.fr,
guy.grunfelder@insa-rennes.fr, ghais.el-zein@insa-rennes

## ABSTRACT
A 60 GHz wireless Gigabit Ethernet (G.E.) communication system is developed at IETR. As the 60 GHz radio link operates only in a single-room configuration, an additional Radio over Fibre (RoF) link is used to ensure the communications in all the rooms of a residential environment. The realized system covers 2 GHz bandwidth. Due to the hardware constraints, a symbol rate at 875 Mbps is attained using simple single carrier architecture. In the baseband (BB) processing block, an original byte/frame synchronization process is designed to provide a smaller value of the preamble missing detection and false alarm probabilities. Bit error rate (BER) measurements have been realized in a large gym for line-of-sight (LOS) conditions. A Tx-Rx distance greater than 30 meters was attained with low BER using high gain antennas and forward error correction RS (255, 239) coding.

## Categories and Subject Descriptors
C.2.1 [**Network Architecture and Design**]: Wireless communication; C.4 [**Performance of Systems**]: Performance attributes.

## 1. INTRODUCTION
60 GHz wireless systems, currently under standardization, are targeting high data rate for wireless personal area networks (WPANs) applications [1-3]. The 60 GHz band, due to a large allocated bandwidth (57-66 GHz) is one of the most promising solutions to achieve data rate of multi-gigabit per second for short distances. Several works have been presented to describe the wireless channel propagation at these frequencies, mm-wave antennas and key system design [2-5]. As described in [1-3], multi-carrier (MC) and single-carrier (SC) schemes are two practical alternatives for high data rate communications in the 60 GHz band. The SC scheme is known to have a lower peak-to-average-power ratio (PAPR) and more resistance to the phase noise than OFDM. In this paper, we propose a hybrid optical/wireless system for indoor Gbps WPANs. The first system application in a point-to-point configuration is the high-speed file transfer. The RoF is partially used in the system. Due to the cost to directly transmit the 60 GHz signals over fiber, it is reasonable to transmit intermediate frequency (IF) signals over the fiber.

As the 60 GHz band suffers high signal propagation loss, a simple solution is to use directional high gain antennas. Fading contributions are minimized by the spatial filtering of the antenna patterns resulting in a higher coherence time. In [4], using directive antennas, the minimum observed coherence time was 32 ms (for people walking with a speed of 1.7 m/s) which is much higher than the lower limit of 1 ms (for omnidirectional antennas). This means that the channel can be estimated once per few thousands of data symbols for Gbps transmission rate. Indeed, for a point-to-point application, the Doppler effect caused by moving objects is not critical in indoor environments. Furthermore, the root mean square (RMS) delay spread depends also on the antenna beamwidth. A large number of publications have shown that directive antennas greatly reduce multipath propagation effects. In [4], using directive antennas, it was observed that the RMS delay spread can be easily reduced to about 1 ns. This means that the inter-symbol interference (ISI) effects are less critical when the throughput is less than 1 Gbps. Then, due to the non linearities of the power amplifier and the high phase noise at 60 GHz, a simple SC architecture using a differential binary phase shift keying (DBPSK) modulation and differential demodulation is used.

The structure of this paper is as follows. Sections 2 and 3 describe the transmitter (Tx) and the receiver (Rx) respectively. In these sections, the IF and radiofrequency (RF) blocks are firstly presented. Then, the baseband blocks and the byte/frame synchronization method are described. In Section 4, measurement results are presented; this section represents the core contribution of this paper. Section V concludes the work.

## 2. TRANSMITER DESIGN
The transmitter system which consists of baseband (BB), IF and RF blocks is shown in Figure 1. A differential encoder is used before the BPSK modulator, as shown in Figure 1, to remove the phase ambiguity at the receiver (knowing that a differential demodulation is used). The signal should contain timing information that allows clock recovery and bit synchronization at the receiver. Thus, scrambling and preamble must be considered at the transmitter.

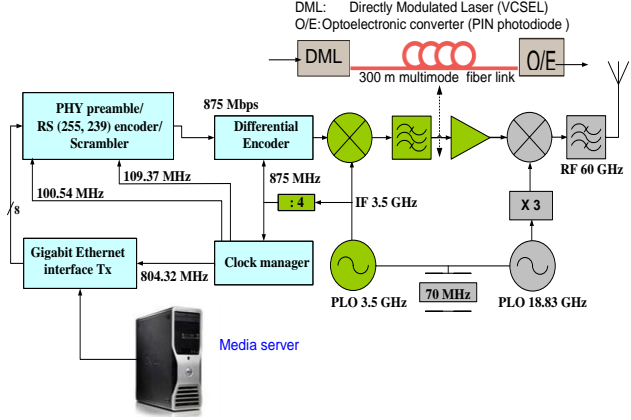

**Figure 1. 60 GHz wireless Gigabit Ethernet (G.E.) Tx**.

## 2.1 Intermediate and Radiofrequency block

After channel coding and scrambling, the input data are differentially encoded using logic components. The differential encoder performs the delayed modulo two addition of the input data with the output. The obtained data are used to modulate an IF carrier generated by a 3.5 GHz phase locked oscillator (PLO) with a 70 MHz external reference. The IF signal is fed into a band-pass filter (BPF) with 2 GHz bandwidth, and transmitted through a 300 meters fibre link. This IF signal is used to modulate directly the current of Laser diode operating at 850 nm. After transmission, the optical signal is converted to an electrical signal by a PIN diode and amplified. The overall RoF link has 0 dB gain.

Following the RoF link, the IF signals is sent to the RF block. This block is composed of a mixer, a frequency tripler, a PLO at 18.83 GHz and a band-pass filter (59-61 GHz). The local oscillator frequency is obtained with an 18.83 GHz PLO with the same 70 MHz reference and a frequency tripler. The phase noise of the 18.83 GHz PLO signal is about –110 dBc/Hz at 10 kHz off-carrier. The BPF prevents spill-over into adjacent channels and removes out-of-band spurious signals caused by the modulator. The 0 dBm obtained signal is fed into the horn antenna with a gain of 22.4 dBi and a half-power beamwidth (HPBW) of 12°.

## 2.2 Baseband architecture

The Tx-G.E. interface is used to connect a home server to a wireless link with about 800 Mbps bit rate, as shown in Figure 2.

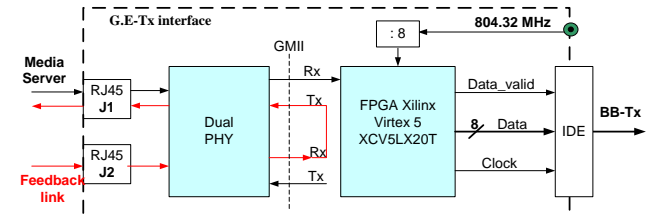

**Figure 2. Transmitter Gigabit Ethernet interface.**

The gigabit media independent interface (GMII) is an interface between the media access control (MAC) device and the physical layer (PHY). The GMII is an 8-bit parallel interface synchronized at a clock frequency of 125 MHz. However, this clock frequency is different from the source byte frequency $f_1 = 804.32/8 = 100.54$ MHz generated by the clock manager in Figure 1. There is risk of packet loss since the source is always faster than the destination. In order to avoid packet losses, a programmable logic circuit is used. The input byte stream is written into the dual port FIFO memory (FPGA) at a high frequency 125 MHz. The FIFO memory has been set up with two thresholds. When the upper threshold is attained, the dual PHY block (controlled by the FPGA) sends a 'signal stop' (to the multimedia source) to stop the byte transfer. The slower frequency $f_1$ reads out continuously the data stored in the register. When the lower threshold is attained, the dual PHY block sends a 'signal start' to launch a new Ethernet frame. Therefore, whatever the activity on the Ethernet access, the throughput at the output of the G.E. interface is constant. Afterward, the byte stream from the G.E. interface is transferred into the dual port FIFO memory, as shown in Figure 3.

A known pseudorandom sequence of 32 bits is completed with one more bit to obtain a 4 bytes preamble. This preamble is sent at the beginning of each frame to achieve good frame synchronization at the receiver. In our application, two kinds of frame structure using 32 bits and 64 bits preamble are considered, as shown in Figure 4 (a) and (b) respectively. An example of 32 bits preamble is only shown in figures of this paper.

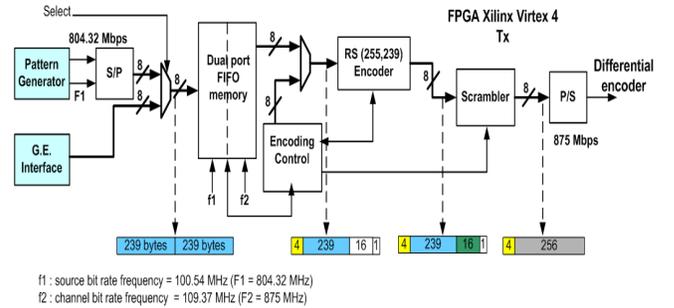

**Figure 3. Transmitter baseband architecture (BB-Tx).**

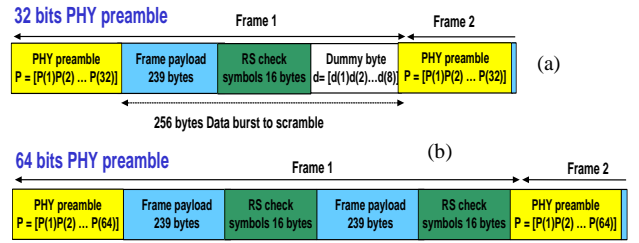

**Figure 4. Frame structure: (a) 32 bits, (b) 64 bits preamble.**

As shown in Figure 4 (a), due to the byte operation of the RS coding, two clock frequencies $f_1$ and $f_2$ are used in the BB block:

$$f_1 = \frac{F_1}{8} = 100.54 \text{ MHz}, \quad f_2 = \frac{F_2}{8} = 109.37 \text{ MHz}$$

where:

$$F_2 = \frac{3.5 \text{ GHz}}{4} = 875 \text{ MHz} \quad \text{and} \quad F_1 = \frac{239}{260} \cdot F_2 = 804.32 \text{ MHz}.$$

As shown in Figure 1, $F_2$ is obtained from the IF.

The frame structure using 32 bits preamble is obtained as follows: the input byte stream is written into the FIFO memory at a slow frequency $f_1$. The dual port FIFO memory has been set up to use two different clock frequencies for writing at $f_1$ and reading at a high frequency $f_2$. Therefore, reading can be started when the FIFO memory is half-full. The encoding control generates 4 bytes preamble and reads 239 bytes of data burst in the register. The RS encoder reads one byte every clock period but bypasses the 4 bytes preamble. After 239 clock periods, the encoding control interrupts the bytes transfer during 17 triggered clock periods, so that check symbol 16 bytes and a dummy byte are added.

This additional dummy byte is only used to complete a number multiple of 4 bytes useful for the scrambling operation. However, this dummy byte is not used in the case of the frame structure using 64 bits preamble, as shown in Figure 4 (b). Therefore, the source bit rate increases slightly at 807.43 Mbps. As shown later by simulations, the used 64 bits preamble can greatly improve the synchronization performance.

Following the RS (255, 239) coding, the byte stream is scrambled using 31 bits PN sequence completed with one more bit to obtain a 4 bytes scrambling sequence. The scrambler bypasses also the 4 bytes preamble in the byte streams. The scrambling sequence is chosen in order to provide at the receiver the lowest false detection of the preamble from the scrambled data. Finally, the byte stream is serialized to output bit stream at 875 Mbps.

## 3. RECEIVER DESIGN

Figure 5 shows the receiver (Rx) system block diagram.

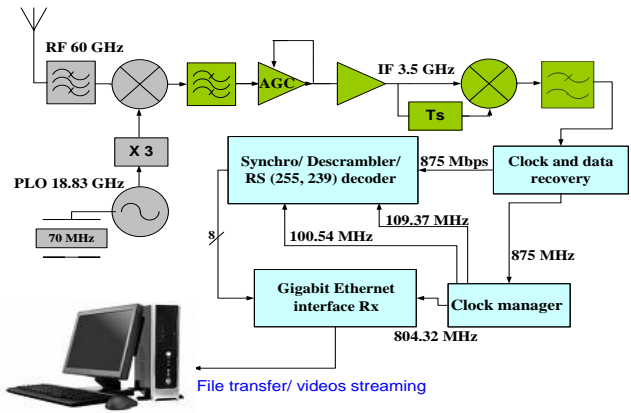

**Figure 5. 60 GHz wireless Gigabit Ethernet Rx.**

### 3.1 Intermediate and radiofrequency block

The receive antenna, identical to the transmit horn antenna, is connected into a band-pass filter (59-61 GHz). The RF filtered signal is down-converted to an IF signal centered at 3.5 GHz and fed into a band-pass filter with a bandwidth of 2 GHz. An automatic gain control (AGC) with a dynamic range of 20 dB is used to ensure a quasi-constant signal level at the demodulator input. A low noise amplifier (LNA) with a gain of 40 dB is used to achieve sufficient gain. A simple differential demodulation enables the coded signal to be demodulated and decoded. In fact, the demodulation, based on a mixer and a delay line (delay equal to the symbol duration Ts = 1.14 ns), compares the signal phase of two consecutive symbols. A "1" is represented as a phase change and a "0" as no change. Due to the product of two consecutive symbols, the rate between the main lobe and the side lobes of the channel impulse response increases. This means that the differential demodulation is more resistant to ISI effect. However, compared to a coherent demodulation, this method is less performing in additive white Gaussian noise (AWGN) channel.

Following the loop, a low-pass filter (LPF) with 1.8 GHz cut-off frequency removes the high-frequency components of the obtained signal. For a reliable clock acquisition realized by the clock and data recovery (CDR) circuit, long sequences of '0' or '1' must be avoided. That's why the use of a scrambler and descrambler is necessary.

### 3.2 Baseband architecture

Figure 6 shows the receiver baseband block diagram. Due to the byte information of the RS (255, 239) decoder, the synchronized data after the CDR are parallelized into byte stream.

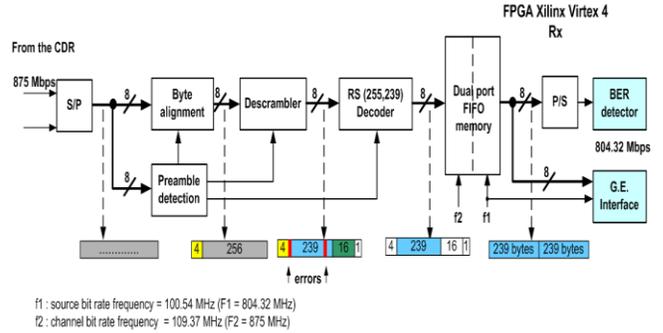

**Figure 6. Receiver baseband architecture (BB-Rx).**

Figure 7 shows the byte/frame synchronization architecture used at the BB-Rx. The frame synchronisation performance is characterized by the miss detection probability ($P_m$), the false alarm probability ($P_F$) and the channel error probability p [6].

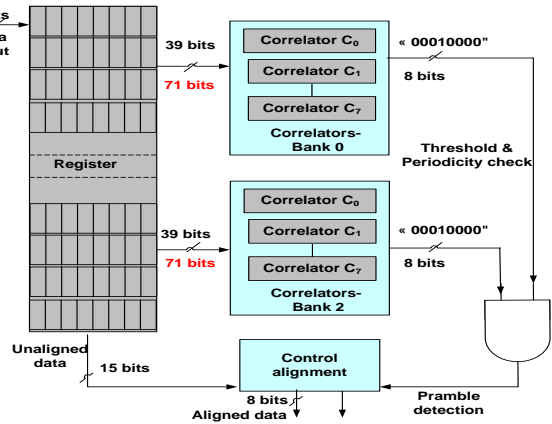

**Figure 7. The preamble detection and byte synchronization.**

The preamble detection is based on the cross-correlation of the 32 bits (or 64 bits) successive received bits and the internal 32 bits (or 64 bits) preamble. Each correlator must analyze a 1-bit shifted sequence. Therefore, the preamble detection is performed with 32+7 = 39 bits (or 64+7 = 71 bits) because of different possible shifts of a byte. In all, due to the different possible shifts of a byte, there are 8 corrrelators in each bank of correlators, as shown in Fig. 7. In order to minimize the preamble missing detection and false alarm probabilities, two banks of correlators are used taking into consideration the periodical repetition of the preamble. Therefore, the decision is made from 264 successive bytes (or 526 bytes) received bits stored by the receiving shift register (P1 + D1 + P2). Each value of the correlation between successive received 32 bits (or 64 bits) and the internal preamble is compared to a threshold $\gamma$. Thereby, the preamble is detected when the same $C_k$ correlator in each bank of correlators indicates its presence.

Figure 8 (a) and (b) show the missing detection probability of the 32 bits and 64 bits preambles, respectively. The threshold setting at the maximum value ($\gamma$ = 32 or 64) is not practical, since a bit error in the preamble leads to a frame loss.

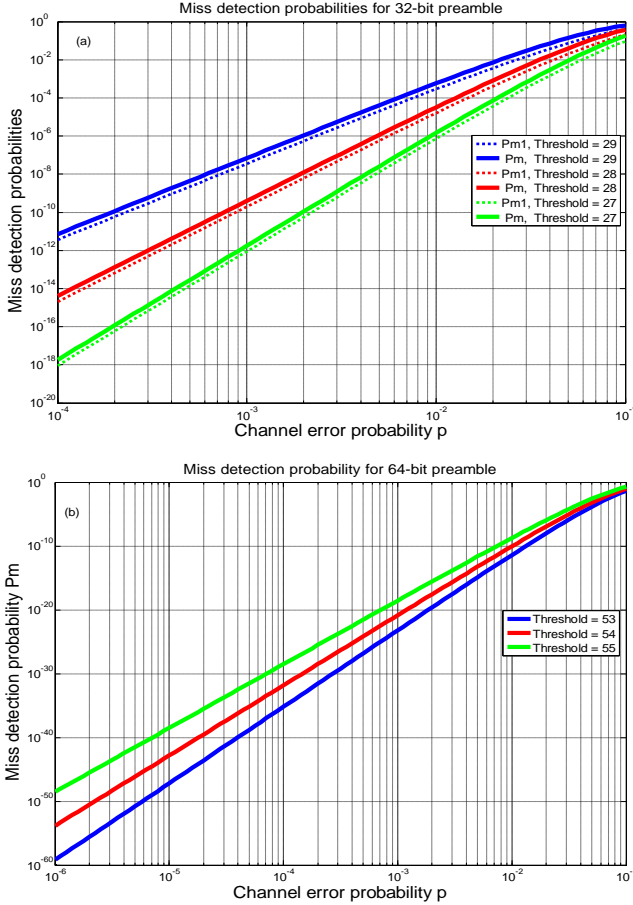
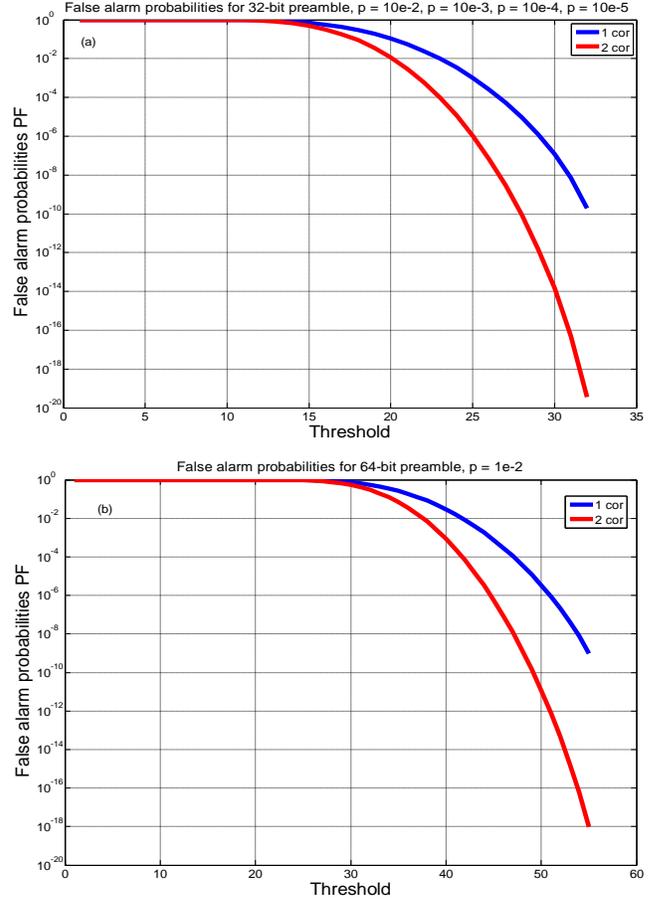

**Figure 8.** $P_m$ of (a) 32 bits and (b) 64 bits preambles.

**Figure 9.** $P_F$ of (a) 32 bits and (b) 64 bits preambles.

The byte/frame synchronization also depends on the false alarm events. A false alarm is declared when the same $C_k$ correlator of each bank of correlators detects the preamble on scrambled data ($D_1$ and $D_2$). Figure 9 (a) and (b) show the false alarm probability with 32 bits and 64 bits preambles, respectively. $P_{F1}$ and $P_{F2}$ indicate the false alarm probability using one and two banks of correlators, respectively. The effect of p on the false alarm probability is insignificant since the random data bits "0" and "1" are assumed to be equiprobable. As a result, the trade-off between $P_F$ and $P_m$ for $p = 10^{-4}$ are obtained as follows:

- 32 bits: $\gamma = 28$, $P_m = 10^{-14}$, $P_{F2} = 10^{-10}$, $P_{F1} = 10^{-5}$,

- 64 bits: $\gamma = 55$, $P_m = 10^{-29}$, $P_{F2} = 10^{-18}$, $P_{F1} = 10^{-6}$.

As the results, for a data rate in the order of 1 Gbps, the byte/frame synchronization can be lost several times using a 32 bits preamble. However, in case of a 64 bits preamble, the trade-off is estimated at $\gamma = 49$, since $P_{F2} = 10^{-20}$. The threshold $\gamma = 49$ provides a smaller value of missing detection and false alarm probabilities of the preamble.

After byte/frame synchronization, the descrambler performs the modulo 2 addition between 4 successive received bytes (or 8 bytes) and the internal 4 bytes (or 8 bytes) descrambling sequence. Then, the baseband processing block tries to regenerate the original bit stream using RS (255, 239) decoding. The RS (255, 239) decoder can correct up to 8 erroneous bytes and operates at a fast clock frequency $f_2$ = 109.37 MHz. After decoding, the byte stream is written discontinuously into the dual port FIFO memory at clock frequency $f_2$. Another slow frequency clock $f_1$ = 100.54 MHz reads out continuously the byte stream stored by the register, since all redundant information is removed. Afterwards, the byte stream is transferred to the receiver Gigabit Ethernet interface. The feedback signal can be transmitted via a wired Ethernet or a Wi-Fi due to its low throughput.

## 4. MEASUREMENT RESULTS

Back-to-back test of the realized system (without RF blocks and AGC loop) was firstly carried out. The aim is to evaluate the BER versus the signal to noise ratio (SNR) at the demodulator input. An external AWGN is added to the IF modulated signal (before the IF-Rx band pass filter). The AWGN is a thermal noise generated and amplified by successive amplifiers. This noise feeds a band pass filter and a variable attenuator. Then, the SNR is varied by changing the AWGN power. Figure 10 (a) and (b) show the spectrum at IF, without and with extra AWGN respectively. Figure 11 shows the BER versus SNR. Thus, for BER = $10^{-5}$, the SNR degradation is about 3.7 and 3 dB for the realized system without and with RS coding respectively. This bandwidth is too wide for a throughput of 875 Mbps. In order to avoid the increased power noise in the band, the filter bandwidth must be reduced to 1.1 GHz (with a roll-off factor $\alpha$ = 0.25). BER measurements have been realized in a large gym of the university campus. The horn antennas Tx-Rx (positioned in the middle of the empty gym) were situated at a height of 1.35 m. These measurements were conducted under LOS conditions with a fixed Rx and the Tx placed on a trolley pushed in the horizontal plane.

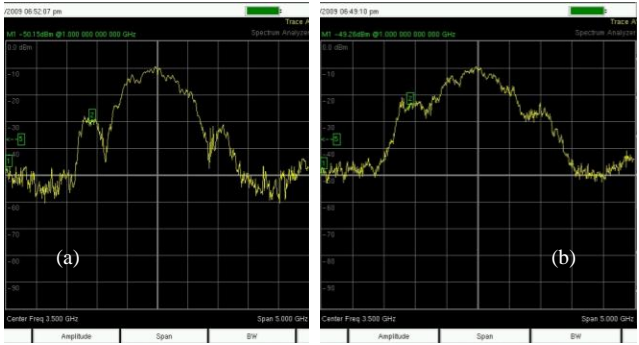

**Figure 10. Spectrum at IF-Rx: (a) without, (b) with AWGN.**

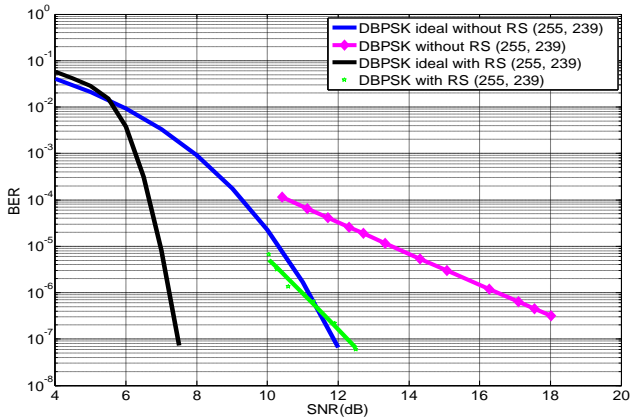

**Figure 11. BER results versus SNR of realized modem.**

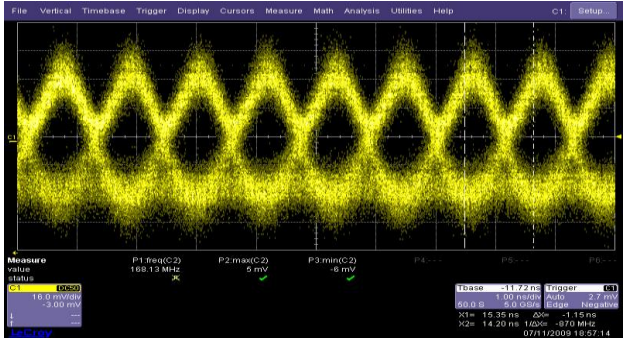

**Figure 12. Eye diagram for a 30 m Tx-Rx distance**.

Figure 12 shows the eye diagram observed for a 30 m Tx-Rx distance. Figure 13 shows results of BER measurements versus Tx-Rx distance, using 32 bits preamble (the BER measurements using 64 bits preamble are under realization). The threshold was set at $\gamma = 29$. The effects of multipath propagation on the BER performance are greatly reduced when directive horn antennas are used. However, for properly aligned antennas, if the direct path is blocked by a moving person, the communication is interrupted (attenuation arround 15 dB [5]). To overcome this problem, it is possible to exploit the angular diversity obtained by switching antennas or by beamforming [3]. A Tx antenna mounted on the ceiling, placed in the middle of the room is also a solution to mitigate the radio beam blockage caused by people or furniture.

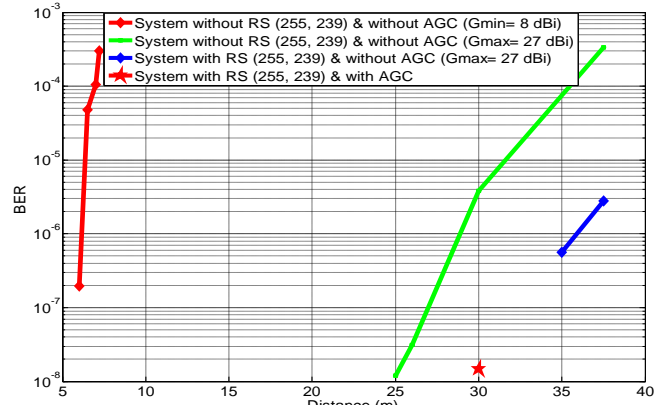

**Figure 13. BER versus Tx-Rx distance (32 bits preamble).**

## 5. CONCLUSION

This paper presents the design and the implementation of a 60 GHz system for multimedia wireless communications. The proposed system provides a good trade-off between performance and complexity. An original method used for the byte synchronization is also described. This method allows a high preamble detection probability and a very small false detection probability. For Gbps reliable transmission within a large room, antennas must have a relatively high gain.

With 2 GHz bandwidth, a data rate of 1.75 Gbps could be transmitted using the same DBPSK modulation. Then, with a new CDR circuit operating up to 2.7 Gbps, a data rate 1.75 Gbps can be achieved. Increasing data rate to 3.5 Gbps is planned with QPSK modulation. Thus, an adaptative equalizer must be implemented to mitigate the ISI effect. The demonstrator will be further enhanced to prove the feasibility of Gbps wireless communications in non line-of-sight environments.

## 6. ACKNOWLEDGMENTS

This work is a part of Techim@ges research project supported by French "Media and Networks Cluster" and Comidom project financed by the "Region Bretagne".